\documentstyle[12pt,a4,epsfig]{article}

\newlength{\abstwidth}
\begin{document}
\newcommand{\am }{{\footnotesize AMPT } }
\newcommand{\amm}{{\footnotesize AMPT}}
\newcommand{\ammm}{{\footnotesize AMPT.} }
\newcommand{\ieh }{{\footnotesize IEH } }
\newcommand{\iehh}{{\footnotesize IEH}}
\newcommand{\flu }{fluctuations }
\newcommand{\fll}{fluctuations}
\newcommand{\cor}{ correlations }
\newcommand{\corr}{correlations}
\newcommand{\et}{ $\eta$ }
\newcommand{\deta}{$\Delta \eta$ }
\newcommand{\ym}{ \(Y_m\) }
\newcommand{\dym}{ \(\frac{\Delta \eta}{Y_m}\) }
\newcommand{\sym}{ \(\frac{S}{Y_m}\) }
\newcommand{\oxy}{ \(^{16}O\) }
\newcommand{\nf}{n$_f$ }
\newcommand{\nb}{n$_b$ }
\thispagestyle{empty}
\begin{center}
\noindent{\Large {\bf Search for Long-Range Correlations in Relativistic Heavy-Ion Collisions at SPS Energies.}}\\[5mm]
\end{center}

\noindent{\bf Shakeel Ahmad,$^a$\footnote{email: Shakeel.Ahmad@cern.ch} Anisa Khatun,$^a$ Shaista Khan,$^a$ A.Ahmad,$^b$ and M.Irfan$^a$}
\begin{enumerate}
\item[a)]{\it  Department of Physics, Aligarh Muslim University,Aligarh 202002,India}\\[-5mm]
\item[b)]{\it  Department of Applied Physics, Aligarh Muslim University,Aligarh 202002, India}\\[-2mm]
\end{enumerate}

\begin{center}
{\bf Abstract}\\[2ex]
\begin{minipage}{\abstwidth} Long range correlations are searched for by analyzing the experimental data on $^{16}$O-AgBr and $^{32}$S-AgBr collisions at 200A GeV/c and the results are compared with the predictions of a multi phase transport(\amm) model. The findings reveal that the observed forward-backward (F-B) multiplicity correlations are mainly of short-range in nature. The range of F-B correlations are observed to extend with increasing projectile mass. The observed extended range of F-B correlations might be due to overall multiplicity fluctuations arising because of nuclear geometry. The findings are not sufficient for making any definite conclusions regarding the presence of long-range correlations.\\

\end{minipage}
\end{center}

\noindent KEY-WORDS: Long-range correlations, Multiparticle production, Relativistic heavy-ion collisions.\\
\phantom{dummy}

\newpage
\noindent {\bf 1. Introduction:}\\
\noindent One of the main goals of studying nucleus-nucleus collisions at relativistic energies is to study the properties of strongly interacting matter under extreme conditions of initial energy density and temperature, where formation of quark-gluon plasma(QGP) is envisaged to take place[1,2,3]. Correlations among the relativistic charged particles produced in different pseudorapidity,{$\eta$} bins are considered as a powerful tool for understanding the underlying mechanism of multiparticle production in hadron-hadron(hh), hadron-nucleus(hA) and nucleus-nucleus(A-A) collisions[4,5,6]. Both short- and long-range correlations have been observed in hadronic and heavy-ion collisions at {\footnotesize SPS} and {\footnotesize RHIC} energies[5,6,7,8,9,10]. These observed correlations have been interpreted in terms of the concept of clustering[11], that is, the particle production takes place via the formation of some intermediate states, referred to as 'clusters' which finally decay isotropically in their centre-of-mass(c.m.) frame to real hadrons. Useful information regarding the properties of clusters, for example, size of clusters, number of clusters produced on event-by-event(ebe) basis and the 'width', the extent of phase space occupied, and so forth, can be extracted by studying the two particle angular correlations[3,12,13]. It has been suggested[4,5,14,15] that inclusive two particle correlations have two components: the short range correlations (SRC) and the long range correlations(LRC). The SRC have been observed to remain confined to a region, $\eta\sim\pm1$ unit around mid rapidity, while the LRC, which arise due to ebe \flu of overall particle multiplicity, extend to a rather longer range[14,15,16] ($>$ 2 units of \et). LRC have been observed at relatively higher incident energies[6,14,15,16,17,18], while the magnitude of LRC, in the case of hh collisions, has been reported to increase with increasing beam energies as the non-singly diffractive inelastic cross-section increases significantly with incident energy for hh collisions at $\sqrt{s}>100$ GeV[19]. These effects have been successfully explained in terms of multiparton interactions[18]. For AA collisions, the multiparton interactions are expected to give rise to LRC, which would extend to rather longer range as compared to those observed in hh collision at the same incident energy[6,15,20,21]. The color glass condensate picture of particle productions and the multiple scattering model also predict presence of LRC in AA collisions[6,8,15,20,22,23].\\
\noindent After the availability of the data from relativistic heavy ion collider({\footnotesize RHIC}) and then from large hadron collider({\footnotesize LHC}), interest in the studies involving particle correlations has considerably increased. It is because of the idea that modifications of the cluster characteristics and (or) shortening in the correlation length in the pseudorapidity space, if observed particularly at these energies may be taken as a signal of transition to quark-gluon plasma formation[4,15,24]. A number of attempts have been made by theoretical and experimental physicists[5,6,13,25,26,27,28,29,30,31,32,33,34,35] to study forward-backward (F-B) correlations at {\footnotesize RHIC} and {\footnotesize LHC} energies. It is, however, essential to identify some baseline contributions to the experimentally observed correlations which do not depend on new physics, for example, formation of some exotic states like {\footnotesize DCC} or {\footnotesize QGP}. It is, therefore, considered worthwhile to carry out a systematic study of F-B correlations at lower energies, {\footnotesize BNL}, and {\footnotesize SPS} because of the fact that only a few attempts have been made to study F-B correlations at these energies[9,14,15,16,36]. Such studies would help understand systematically the underlying physics at energies from {\footnotesize SPS} to {\footnotesize RHIC}, like dependence of correlation strength and correlation length on beam energy and system size. Once such dependence is understood, modification in the cluster characteristics or shortening  of correlation length may be looked into to search for {\footnotesize QGP} formation. \\

\noindent {\bf 2. Formalism:}\\
\noindent F-B correlations are generally investigated by examining the following type of linear dependence of mean charged particle multiplicity in the backward(B) hemisphere, $<n_{b}>$ on the multiplicity of the particles emitted in the forward (F) hemisphere, $n_{f}$:
\begin{equation}
<n_{b}> = a + bn_{f}
\end {equation}
\noindent where a is intercept and b represents the slope. For symmetric F and B regions, b is often termed as the correlation strength and is expressed in terms of expectation value[6,15,28,37]:
\begin{equation}
b=\frac{<n_{f}n_{b}>-<n_{b}><n_{f}>}{<{n^{2}_{f}}> - {<n_{f}>}^{2}}=\frac{D_{bf}^{2}}{D_{ff}^{2}}\\
\end{equation}
\noindent where $D_{ff}$ and $D_{bf}$ denote the forward-forward and backward-forward dispersions, respectively.\\


\noindent{\bf 3. Details of the data }\\ 
\noindent Two samples of events, produced in the interactions of \(^{16}\)O and \(^{32}\)S ions with AgBr group of nuclei in emulsion at 200A GeV/c are used in the present study; the number of events produced in \(^{16}\)O-AgBr and \(^{32}\)S-AgBr interactions are 223 and 452 respectively. These events are taken from the collection of emulsion experiments performed by EMU01 collaboration[38]. The other relevant details of the data, like, criteria for selection of events, classification of tracks, selection of AgBr group of events, and so forth, may be found elsewhere[4,38,39,40]. The emission angle, $\theta$ of the relativistic charged particle with respect to beam axis were measured by the coordinate method. The values of x, y, z coordinates at the vertex and at two points one on shower and the other on beam tracks were measured and the pseudorapidity variable, $\eta$ was calculated using the relation, \(\eta = -lntan(\theta/2)\). It should be emphasized that the conventional emulsion technique has two main advantages over the other detectors: (i) its 4{$\pi$} solid angle coverage and (ii) emulsion data are free from biases due to full phase space coverage. In the case of other detectors, only a fraction of charged particles are recorded due to the limited acceptance cone. This not only reduces the charged particle multiplicity but may also distort some of the events characteristics, such as particle density fluctuations[4,41]. In order to compare the findings of the present work with a multi phase transport model,{\footnotesize AMPT}[42], two samples of events corresponding to \(^{16}\)O-AgBr and \(^{32}\)S-AgBr collisions at 200A GeV/c are simulated using the Monte Carlo code, ampt-v1.21-v2.21; the number of events in each sample is equal to that in the experimental data sample. The events are simulated by taking into account the percentage of interactions which occur in the collisions of projectile with various target nuclei in emulsion[43,44]. The values of impact parameter for each data set is so set that the mean multiplicities of relativistic charged particles becomes nearly equal to those obtained for the experimental data sets.\\
\noindent The \am model is a mixed model based on both hadronic and partonic phases[44]. There are four subprocesses in this model[44,45]; phase space initialization, the parton-parton interactions, the conversion from partonic to the hadronic matter and the late hadronic interactions. The initialization takes the {\footnotesize HIJING} model[46] as event generator which included minijet production and soft string excitation. Scattering among the partons follows Zhang's Parton Cascade ({\footnotesize ZPC}) model[47]. The hadronization process is described by Quark Coalescence Model[44] in which two nearest partons combine to become a meson and three nearest partons combine to form a baryon. Finally the rescattering and resonance decay of partons are described by {\footnotesize ART} (a relativistic transport) model[48].\\
\noindent Pseudorapidity distribution of relativistic charged particles for the experimental and \am event samples at the two incident energies considered are displayed in Figure 1. It is interesting to note in the figure that the distributions corresponding to experimental and \am events acquire almost similar shapes.\\


\noindent{\bf 4. Results and discussion}\\
\noindent Pseudorapidity, {$\eta$} distribution of relativistic charged particles is divided into two parts with respect to its center of symmetry, ${\eta}_{c}$. The region with values ${\eta}<{\eta}_{c}$ is referred to  as the backward (B) region while the region having values ${\eta}>{\eta}_{c}$ is termed as the forward (F) region. The number of relativistic charged particles emitted in F and B regions are counted on event-by-event (ebe) basis and hence the mean multiplicities in the two regions, $<n_{f}>$ and $<n_{b}>$ and dispersions $D_{ff}$ and $D_{bf}$ are estimated. Dependence of $<n_{b}>$ on $n_{f}$ for various data sets considered are displayed in Figure 2. The straight lines in the figure represent the best fit to data obtained using (1). The values of slope parameter b obtained from the linear fits are listed in Table 1. Values of b for various data sets are also calculated using (2) and are listed in Table 1. It may be noted from the table that values of b obtained from the linear fits are nearly equal to the corresponding values estimated using (2). F-B correlation strength, thus estimated from either (1) or (2), indicates the presence of F-B correlations in both experimental and simulated data sample. It may also be noted from Table 1 that the values of correlation strength b are nearly the same for \(^{16}\)O and \(^{32}\)S-AgBr collisions. However, for \(^{16}\)O-AgBr collisions at 14.5,60,and 200A GeV/c values of b have been observed[15] to decrease with increasing beam energy. This indicates that correlation strength in the case of AA collisions decreases with increasing incident energy but remains nearly constant with increasing projectile mass. The larger values of b at lower energies observed in \(^{16}\)O-AgBr collisions might be due to the dominance of uncorrelated production for which F-B correlations depend on the mean multiplicity and multiplicity fluctuations in the combined F-B regions[14,15,16,34].\\
\noindent Strong F-B correlations are observed when F and B regions are selected such that there is no separation gap between the two regions. This may be attributed mainly to the clusters produced around ${\eta}_{c}$ whose decay product would go to both F-and B-regions, giving rise to strong SRC. The SRC are envisaged to be confined to a region of $\pm1{\eta}$ units around ${\eta}_{c}$[14,15,16,34]. In order to minimize the contributions from SRC, a gap of ${\Delta}{\eta}$ from the center of symmetry is introduced in both F-and B-regions such that the particles having {$\eta$} values ${\eta}_{c}<{\eta}<{\eta}_{c}+{\Delta}{\eta}$ in F-region and ${\eta}_{c}>{\eta}>{\eta}_{c}-{\Delta}{\eta}$ in B-region are not considered while evaluating ${n}_{f}$ and ${n}_{b}$. The values of correlation strength b are then calculated by estimating ${D_{ff}^{2}}$ and ${D_{bf}^{2}}$ (using 2) by taking ${\Delta}{\eta}$=0.25 and then increasing its value in step of 0.25. The variation of b, with ${\Delta}{\eta}$ thus obtained for the experimental and \am data sets are  plotted in Figure 3. It is observed that values of b, for both \(^{16}\)O-AgBr and \(^{32}\)S-AgBr collisions, remain essentially constant upto ${\Delta}{\eta} {\simeq} $1.0 and thereafter gradually decrease to 0 with increasing ${\Delta}{\eta}$. \am data too exhibit a similar trend of variations of b with \deta. It may however be noted that \am predicts somewhat smaller values of b in the region of smaller ${\Delta}{\eta}$ (${\Delta}{\eta}<$1.25) and relatively larger values of b in the region of \deta $\ge$ 1.5. The smaller values of b observed for \am data as compared  to the corresponding experimental data in the region \deta $\le$ 1.25 might be due to the dominance of uncorrelated production in the \am model; the exact cause of uncorrelated production in the \am model could not be ascertained. Beyond this region, that is,\deta $\ge$ 1.5, values of b  are noticed to be larger for \am events as compared to those obtained from the experimental data. \am thus gives a slower decrease in the values of b with \deta in comparison to that observed with experimental data. Thus, in the case of \am events F-B correlations are observed to characteristically extend to rather longer range as compared to those observed with the experimental data. Furthermore, almost similar values of b, for both \(^{16}\)O and \(^{32}\)S projectiles, as is evident from Figure 4, indicate that the correlation strength is independent of the mass of the colliding beam. This observation is well supported by the \am model. Some difference in the b values for \(^{16}\)O-AgBr and \(^{32}\)S-AgBr experimental events in the region \deta $\sim$ 2.0 might be because of the fluctuations arising due to limited statistics.\\
\noindent It has been reported [15,34] that multiplicity distributions have different shapes in different pseudorapidity regions and exhibit large fluctuations in wider $\eta$-windows. In order to examine the F-B correlation strength in $\eta$ windows of different widths, two small windows each of width \(\eta_w\) = 0.25 are placed adjacent to each other with respect to \(\eta_c\) such that the charged particles having $\eta$ values in the range \(\eta_c \leq\eta < \eta_c + \eta_w\) are counted as $n_{f}$ while those having their $\eta$ values lying in the interval \(\eta_c > \eta \ge \eta_c - \eta_w\) are counted as $n_{b}$ and the value of correlation strength, b is computed. The width, \(\eta_w\) is then increased in step of 0.25 until almost entire $\eta$ region is covered. Variations of b with \(\eta_w\) for the experimental and \am data are shown in Figure 5. It may be noted from the figure that the values of b first increases slowly with increasing \(\eta_w\) (upto \( \eta_w \sim\)2.0) and thereafter acquires nearly constant values. Similar trends of variations of b with \(\eta_w\) have also been observed  earlier for 14.5A, 60A and 200A GeV/c $^{16}$O-AgBr collisions[9,15]. It may also be noted from the figure that although \am predicts the similar trends of variations of b with \(\eta_w\) for both the data sets yet it is evidently clear that \am predicted values are somewhat smaller as compared to those observed for the experimental data in the entire range of \(\eta_w\) considered. Furthermore, it is also clear from Figure 5 that the values of b for any given \(\eta_w\) are nearly the same for both the data sets. This suggests that the values of  b are independent of the mass of the colliding nuclei. It should be mentioned here that, in the saturation region, that is, the region,(${\eta_{w}}>$ 1.5), values of b, for the experimental data have been reported[15] to decrease with increasing projectile energy. Such a decrease in the values of b has been observed due to the increase in the ratio  \(<n_f>/<n_s>\) even in the limited phase space[15]; \(<n_f>\) denotes the average number of charged particles in the F region and while \(<n_s>\) is the mean charged particle multiplicity in the considered phase space.\\
\noindent In order to examine the presence of LRC, if any, contribution from SRC is to be eliminated. For this purpose F-B correlations are studied by adopting the method which has frequently been used, particularly at {\footnotesize RHIC} and {\footnotesize LHC} energies[5,6,25,27,28,29,30,31,32,34]. According to this method, $\eta$ windows of small but equal widths, \(\eta_w\) are placed in F and B regions in such a way that they are separated by equal distances(in $\eta$ units), \(\eta_{gap}\) with respect to centre of symmetry \(\eta_c\). Thus, all the charged particles having their $\eta$ values in the interval \(\eta_c+\eta_{gap} \leq \eta < \eta_c+\eta_{gap}+ \eta_w\) are counted as $n_{f}$ where as those having their $\eta$ values in the range \(\eta_c-\eta_{gap} \leq \eta < \eta_c-\eta_{gap}-\eta_w\) are counted as $n_{b}$. By changing the value of \(\eta_{gap}\) from 0 to 3.0 on each side of \(\eta_c\), $n_{f}$ and  $n_{b}$ are estimated to evaluate the values of b. Variations of b with $\eta_{gap}$ for various data sets considered are displayed in Figure 6.  It may be noted in the figure that the values of b acquire almost constant value of $\sim$ 0.7 upto  \( \eta_{gap}\sim\)1.25 for $^{16}$O-beam and thereafter suddenly decreases to zero with increasing $\eta$ gap values. For $^{32}$S-beam the values of b are observed to remain constant upto \( \eta_{gap}\sim\)1.75 and then decreases to zero. This indicates that with increasing projectile mass the F-B correlations extend to rather longer range. \am data, too, exhibit similar trends of variations of b with \(\eta_{gap}\) except that the values of  b are somewhat smaller in comparison to the one obtained for the experimental data. These observed correlation are envisaged to be due to formation of resonance or clusters in the central rapidity region, the decay products of which would be emitted in both F and B regions[11,14,16,17]. This observation is not sufficient to consider it as an indication of the presence of some LRC but it does suggest that the range of F-B correlations extends with increasing mass of the projectile. The range of F-B correlations has also been observed to increase with increasing beam energy in $^{16}$O-AgBr collisions in the energy range from 14.5A to 200A GeV/c[15]. It has been argued[34] that the extended range of F-B correlations may be explained from simple statistical considerations of uncorrelated production of charged particles. Correlations in this range, if observed at higher beam energy or with heavier projectile, arise due to overall multiplicity fluctuations[6,14,16,17,34]; such fluctuations in AA collisions may show-up because of fluctuations in nuclear geometry[34]. It has also been pointed out[34] that before drawing up any conclusions regarding the presence of dynamical LRC, it should be confirmed  that the observed F-B correlations are not arising due to overall multiplicity fluctuations by studying the multiplicity distributions and F-B correlations simultaneously in the same experiment.          

\noindent{\bf 5. Summary}\\
\noindent On the basis of the findings of the present work, the following conclusions may be arrived at:
\begin{enumerate}
\item The observed F-B correlations are mainly of short-range in nature. However, the range of F-B correlations are observed to increase with increasing projectile mass and beam energy. This extended range of correlations at higher beam energy or larger projectile mass may be due to overall multiplicity fluctuations arising because of nuclear geometry.
\item The study of F-B correlations dependences on the pseudorapidity bin-width and position indicates that the correlation strength b remains independent of the projectile mass.
\item The Monte Carlo model, \am is observed to reproduce the data nicely.    
\end{enumerate}   
\noindent  {{\bf References}}
\begin{enumerate}
\item[1.] S. Ahmad, A. Ahmad, A. Chandra, M. Zafar, and M. Irfan,
"Entropy analysis in relativistic heavy-ion collisions," {\it Advances
in High Energy Physics}, vol. 2013, Article ID 836071, 10 pages, 2013.
\item[2.] M. Rybczynski, "STAR's measurement of correlations and
fluctuations between photons and charged particles at forward
rapidities at RHIC," {\it Journal of Physics G}, vol. 35, no. 10, Article ID 104094, 2008.)
\item[3.] E. V. Shuryak, "Quantum chromodynamics and the theory of
superdense matter," {\it Physics Reports. Section C of Physics Letters}, vol. 61, no. 2, pp. 71–158, 1980.
\item[4.] S. Ahmad, A. Ahmad, A. Chandra et al., "Forward-backward
multiplicity fluctuations in 200A GeV/c $^{16}$O-AgBr and $^{32}$S-
AgBr collisions," {\it Physica Scripta}, vol. 87, no. 4, Article ID 045201,
2013.
\item[5.] B. Alver, B. B. Back, M. D. Baker et al., "Cluster properties from
two-particle angular correlations in p+p collisions at $\sqrt{s}=200$
and 410 GeV," {\it Physical Review C}, vol. 75, Article ID 054913, 2007.
\item[6.] B. I. Ablev, M. M. Aggarwal, Z. Ahammed et al., "Growth
of long range forward-backward multiplicity correlations with
centrality in Au+Au collisions at $\sqrt{s}_{NN}=200$ GeV," {\it Physical
Review Letters}, vol. 103, no. 17, Article ID 172301, 2009.
\item[7.] T. J. Tarnowsky, "Probing the quark-gluon phase transition
using energy and system-size dependence of long-range mul-
tiplicity correlations in heavy ion collisions from the STAR
experiment," {\it Indian Journal of Physics}, vol. 85, no. 7, pp. 1091–
1095, 2011.
\item[8.]B. K. Srivastava, R. P. Scharenberg, and T. J. Tarnowsky, "Under-
standing the particle production mechanism with correlation
studies using long and short range correlations," {\it International
Journal of Modern Physics E}, vol. 16, p. 2210, 2007.
\item[9.] G. Singh, K. Sengupta, A. Z. M. Ismail, and P. L. Jain, "Long-
range correlations in nucleus-nucleus interactions at ultrahigh
energies," {\it Physical Review C}, vol. 39, no. 5, pp. 1835–1839, 1989.
\item[10.] A. Shakeel, W. B. Tak, N. Ahmad et al., "Cluster production in
14.5 A GeV/c Si-NUCLEUS collisions," {\it International Journal of
Modern Physics E}, vol. 8, no. 2, p. 121, 1999.
\item[11.] E. L. Berger, "Rapidity correlations at fixed multiplicity in
cluster emission models," {\it Nuclear Physics B}, vol. 85, no. 1, pp.
61-101, 1975.
\item[12.] W. Li, "Two-particle angular correlations in p+p and Cu+Cu
collisions at PHOBOS," {\it Journal of Physics G: Nuclear and
Particle Physics}, vol. 34, no. 8, Article ID S1005, 2007.
\item[13.] W. Li, "System size dependence of two-particle angular correlations in p+p, Cu+Cu and Au+Au collisions," {\it Journal of Physics
G}, vol. 35, no. 10, Article ID 104142, 2008.
\item[14.] R. E. Ansorge, B. Asman, C. N. Booth et al., "Charged particle correlations in pp collisions at c.m. energies of 200, 546 and 900GeV," {\it Zeitschrift f$\ddot{u}$r  Physik C}, vol. 37, no. 2, pp. 191–213, 1988.
\item[15.] S. Ahmad, A. Chandra, M. Zafar, M. Irfan, and A. Ahmad, "Short- and long-range multiplicity correlations in relativistic heavy-ion collisions," {\it International Journal of Modern Physics E}, vol. 22, no. 9, Article ID 1350066, 2013.
\item[16.] K. Alpgard, S. Berglunde, K. Berkelman et al., "Forward-
backward multiplicity correlations in $\bar{p}p$-collisions at $\sqrt{s}$=540 Gev," Physics Letters B, vol. 123, no. 5, pp. 361–366, 1983.
\item[17.] G. J. Alner, K. Alpgard, P. Anderer et al., "UA5: a general study
of proton-antiproton physics at $\sqrt{s}=546$ GeV," {\it Physics Reports},
vol. 154, no. 5-6, pp. 247–283, 1987.
\item[18.]T. Alexopoulos, C. Allen, E. W. Anderson et al., "Charged
particle multiplicity correlations in p$\bar{p}$ collisions at $\sqrt{s}=$0.3-1.8 TeV," {\it Physics Letters B}, vol. 353, no. 1, pp. 155–160, 1995.
\item[19.] W. D. Walker, "Multiparton interactions and hadron structure,"
{\it Physical Review D}, vol. 69, Article ID 034007, 2007.
\item[20.] A. Capella, U. Sukhatme, C.-I. Tan, and J. Tran Thanh Van,
"Dual parton model," {\it Physics Reports}, vol. 236, no. 4-5, pp. 225–
329, 1994.
\item[21.] Y. V. Kovchegov, E. Levin, and L. McLerran, "Large scale
rapidity correlations in heavy ion collisions," {\it Physical Review C},
vol. 63, Article ID 024903, 2001.
\item[22.] B. K. Srivastava, "Fluctuations and correlations in STAR," {\it The
European Physical Journal A}, vol. 31, no. 4, pp. 862–867, 2007.
\item[23.] A. Bzdak, "Forward-backward multiplicity correlations in
AuAu collisions," {\it Acta Physica Polonica B}, vol. 40, no. 7, pp.
2029–2032, 2009.
\item[24.] V. P. Konchakovski, M. Hauer, G. Torrieri, M. I. Gorenstein, and
E. L. Bratkovskaya, "Forward-backward correlations in nucleus-
nucleus collisions: baseline contributions from geometrical
fluctuations," {\it Physical Review C}, vol. 79, Article ID 034910, 2009.
\item[25.] B. Alver, B. B. Back, M. D. Bake et al., "System size dependence
of cluster properties from two-particle angular correlations in
Cu + Cu and Au + Au collisions at $\sqrt{s}_{NN}=$200 GeV," {\it Physical
Review C}, vol. 81, Article ID 024904, 2010.
\item[26.] J. L. Albacete, A. Dumitru, and C. Marquet, "The initial state of
heavy-ion collisions," {\it International Journal of Modern Physics A},
vol. 28, Article ID 1340010, 2013.
\item[27.] M. Skoby, "Forward-backward multiplicity correlations at
STAR," {\it Nuclear Physics A}, vol. 854, no. 1, pp. 113–116, 2011.
\item[28.] T. J. Tarnowski, "Recent results of fluctuation and correlation
studies from the STAR experiment," {\it Journal of Physics: Conference Series}, vol. 230, no. 1, Article ID 012025, 2012.
\item[29.] B. Alver, B. B. Back, M. D. Baker et al., "Multiplicity fluctuations
in Au + Au collisions at RHIC," {\it International Journal of Modern
Physics E}, vol. 16, no. 7-8, pp. 2187–2192, 2007.
\item[30.] A. K. Dash, D. P. Mahapatra, and B. Mohanty, "Expectation of
forward-backward rapidity correlations in p+p collisions at the
LHC energies," {\it International Journal of Modern Physics A}, vol.
27, no. 14, Article ID 1250079, 2012.
\item[31.] Y. L. Yan, D.-M. Zhou, B.-G. Dong et al., "Centrality dependence
of forward-backward multiplicity correlation in Au+Au colli-
sions at $\sqrt{s}_{NN}=$ 200 GeV," {\it Physical Review C}, vol. 81, Article ID
044914, 2010.
\item[32.] Y. L. Yan, B. G. Dong, D. M. Zhou, X. M. Li, H. L. Ma, and B. H.
Sa, "Forward-backward multiplicity correlations in pp, pp and
Au+Au collisions at RHIC energy," {\it Nuclear Physics A}, vol. 834,
pp. 320C-322C, 2010. (2010).
\item[33.] A. Abdelsalam, M. S. El-Nagdy, and E. A. Shaat, "Study of
relativistic forward-backward hadron production in the inter-
actions of 3He and 4He with emulsion nuclei at Dubna energy,"
{\it Fizika B}, vol. 15, no. 1, pp. 9-22, 2006.
\item[34.] J. Fu, "Statistical interpretation of multiplicity distributions and
forward-backward multiplicity correlations in relativistic heavy
ion collisions," {\it Physical Review C}, vol. 77, Article ID 027902,
2008.
\item[35.] A. Korner and M. Lublinsky, "Angular and long range rapidity
correlations in particle production at high energy," {\it International
Journal of Modern Physics E}, vol. 22, no. 1, Article ID 1330001,
2013.
\item[36.] P. L. Jain, K. Sengupta, and G. Singh, "Short- and long-
range correlations of produced particles at very high energies,"
{\it Physical Review D}, vol. 34, pp. 2886-2889, 1986.
\item[37.] A. Capella and A. Krzywicki, "Unitarity corrections to short-
range order: long-range rapidity correlations," {\it Physical Review
D}, vol. 18, p. 4120, 1978.
\item[38.]M. I. Adamovich, M. M. Aggarwal, Y. A. Alexandrov et al.,
"Produced particle multiplicity dependence on centrality in
nucleus-nucleus collisions," {\it Journal of Physics G}, vol. 22, no.
10, p. 1469, 1996.
\item[39.] S. Ahmad, M. M. Khan, N. Ahmad, and A. Ahmad, "Erraticity
behaviour in relativistic nucleus-nucleus collisions," {\it Journal of
Physics G}, vol. 30, no. 9, pp. 1145–1152, 2004. 
\item[40.] S. Ahmad, A. R. Khan, M. Zafar, and M. Irfan, "On multifractality and multifractal specific heat in ion-ion collisions," {\it Chaos,
Solitons and Fractals}, vol. 42, no. 1, pp. 538-547, 2009.
\item[41.] M. L. Cherry, A. Dabrowska, P. Deines-Jones et al., "Event-by-event analysis of high multiplicity Pb(158 GeV/nucleon)-Ag/Br
collisions," {\it Acta Physica Polonica B}, vol. 29, no. 8, pp. 2129-2146,
1998.
\item[42.] Z. W. Lin, C. M. Ko, B. A. Li, B. Zhang, and S. Pal, "Multiphase
transport model for relativistic heavy ion collisions," {\it Physical
Review C}, vol. 72, Article ID 064901, 2005.
\item[43.] C. F. Powell, P. H. Fowler, and D. H. Perkins, {\it The Study of
Elementary Particles by Photographic Method}, Pergamon Press,
Oxter, UK, 1959.
\item[44.] M. I. Adamovich, Y. A. Alexandrov, S. A. Asimov et al., "Multiplicities and rapidity densities in 200 A GeV $^{16}$O interactions
with emulsion nuclei," {\it Physics Letters B}, vol. 201, no. 3, pp. 397-
402, 1988. 
\item[45.] Y.-L. Xie, G. Chen, J.-L. Wang, Z.-H. Liu, and M.-J. Wang,
"Scaling properties of multiplicity fluctuations in heavy-ion
collisions simulated by AMPT model," {\it Nuclear Physics A}, vol.
920, pp. 33-44, 2013.
\item[46.] N. Li, S. Shi, A. Tang, and Y. Wu, "The study of non-collectivity
by the forward–backward multiplicity correlation function,"
{\it Journal of Physics G}, vol. 39, no. 11, Article ID 115105, 2012.
\item[47.] X. N. Wang, "Role of multiple minijets in high-energy hadronic
reactions," {\it Physical Review D}, vol. 43, p. 104, 1991.
\item[48.] B. Zhang, "ZPC 1.0.1: a parton cascade for ultrarelativistic heavy
ion collisions," {\it Computer Physics Communications}, vol. 109, no.
2-3, pp. 193-206, 1998.
\end{enumerate}

\newpage
\noindent Table~1:  Values of correlation strength, \(b \) and \(\chi^2/D.F. \) for the experimental and \am event samples at different  projectile energies.
\begin{center}
\begin{tabular}{lcccc} \hline 
Energy        & \multicolumn{2}{c}{b (linear fit)} & \multicolumn{2}{c}{b(=\(\frac{D^2_{bf}}{D^2_{ff}}\))} \\ [1mm]
 (GeV)      & Expt. & \am & Expt. & \am \\ \hline
        $^{16}$O-AgBr  & 1.21$\pm$0.06 &  1.08$\pm$0.06 & 1.20$\pm$0.03 & 1.07$\pm$0.05 \\[2mm] 
 $^{32}$S-AgBr  & 1.19$\pm$0.03 &  1.03$\pm$0.02 & 1.18$\pm$0.02 & 1.03$\pm$0.03 \\ \hline

\end{tabular}

\end{center}
 
\newpage
\begin{figure}[]
\begin{center}\mbox{\psfig{file=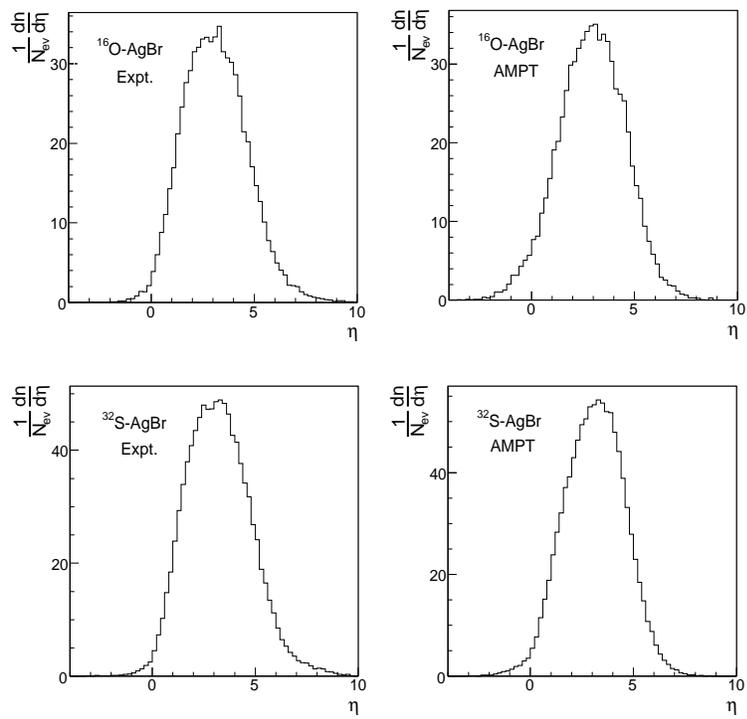,width=10cm,height=10cm}}
\end{center}
\caption[Fig1.]{\sf{Pseudorapidity distributions of relativistic charged particles produced in $^{16}$O- and $^{32}$S-AgBr collisions compared with \am predictions.}}
\label{lindat}
\end{figure}

\newpage
\begin{figure}[]
\begin{center}\mbox{\psfig{file=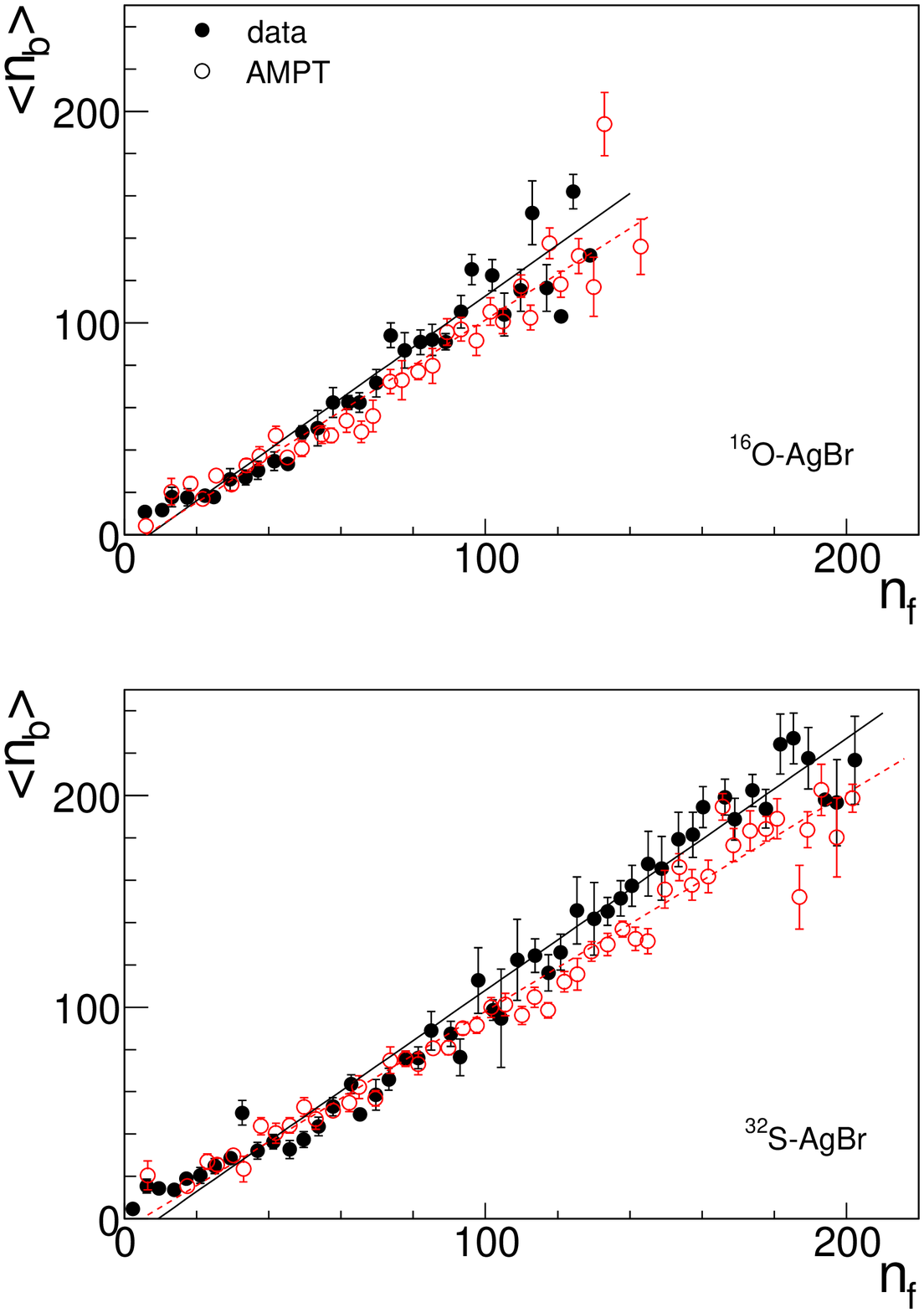,width=10cm,height=16cm}}
\end{center}
\caption[Fig1.]{\sf{Variations of \(<n_b>\) with n$_f$ for $^{16}$O- and $^{32}$S-AgBr collisions. The straight lines represent the best fit to the data obtained using Eq.(1).}}
\label{lindat}
\end{figure} 

\newpage
\begin{figure}[]
\begin{center}\mbox{\psfig{file=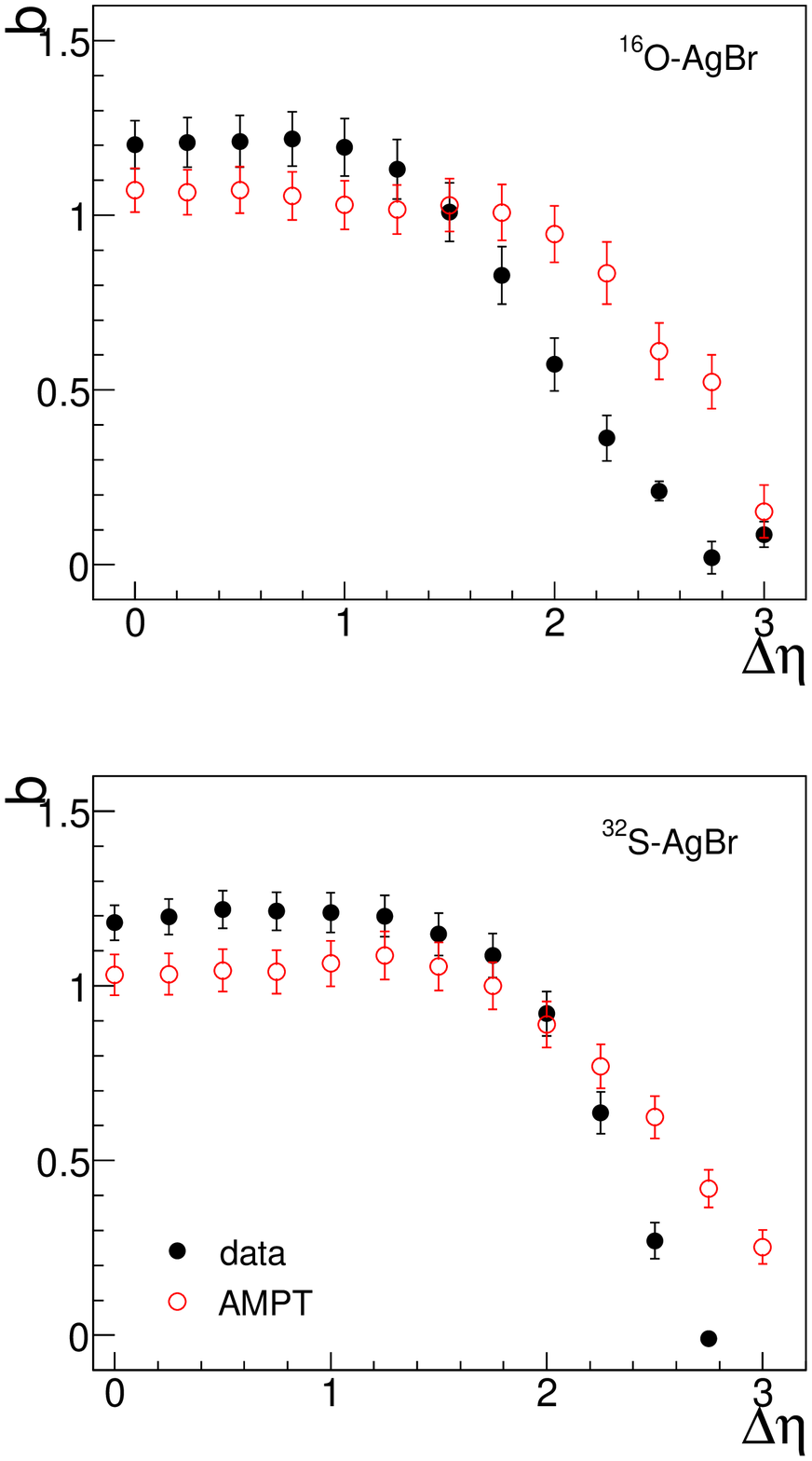,width=10cm,height=16cm}}\end{center}
\caption{\sf Variations of correlation strength b with pseudorapidity window width, \deta for $^{16}$O- and $^{32}$S-AgBr collisions. }
\label{lindat}
\end{figure}

\newpage
\begin{figure}[]
\begin{center}\mbox{\psfig{file=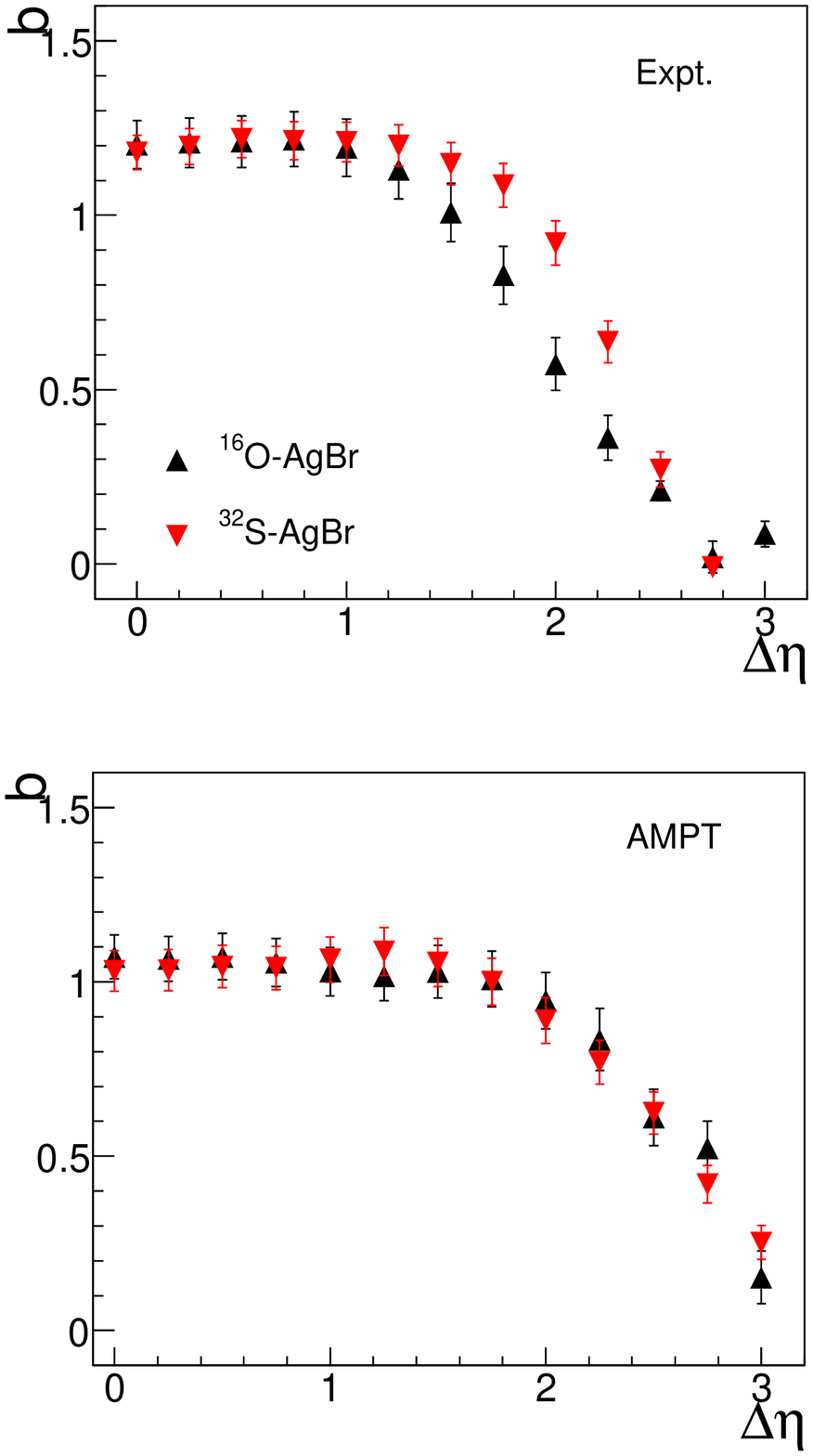,width=10cm,height=16cm}}\end{center}
\caption{\sf Variations of b with \deta for $^{16}$O- and $^{32}$S-AgBr collisions. }
\label{lindat}
\end{figure}

\newpage
\begin{figure}[hptb]
\begin{center}\mbox{\psfig{file=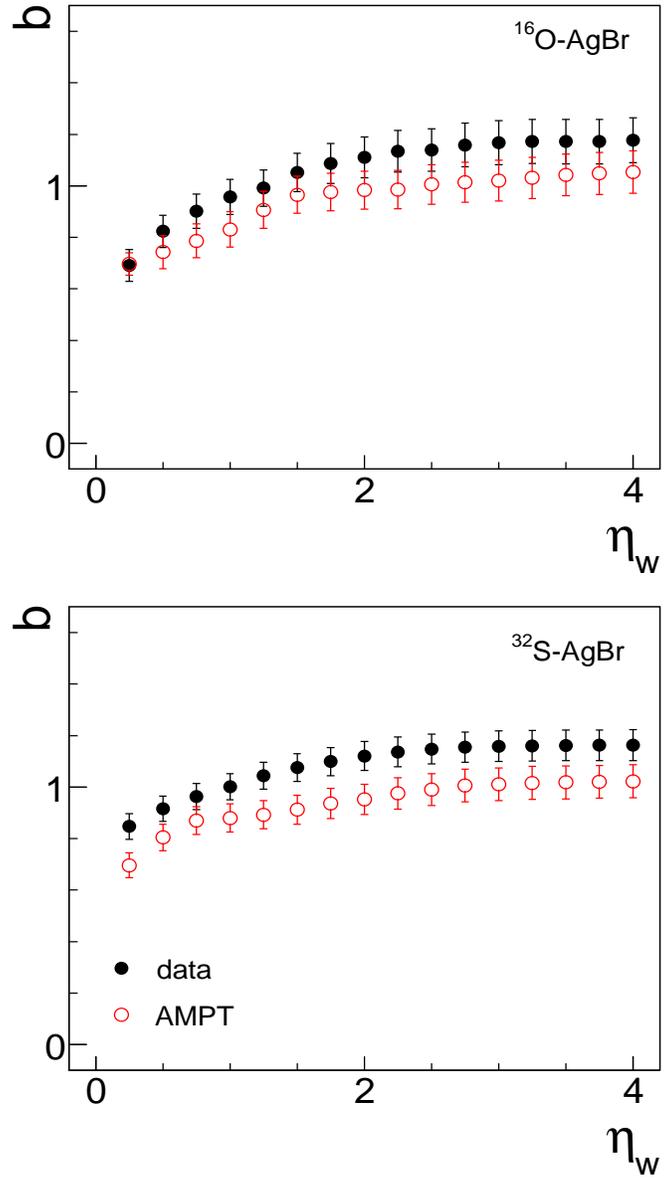,width=10cm,height=16cm}}\end{center}
\caption{{\sf Dependence of b on separation gap, \(\eta_w\) for $^{16}$O- and $^{32}$S-AgBr collisions.}}
\label{lindat}
\end{figure}

\newpage
\begin{figure}[hptb]
\begin{center}\mbox{\psfig{file=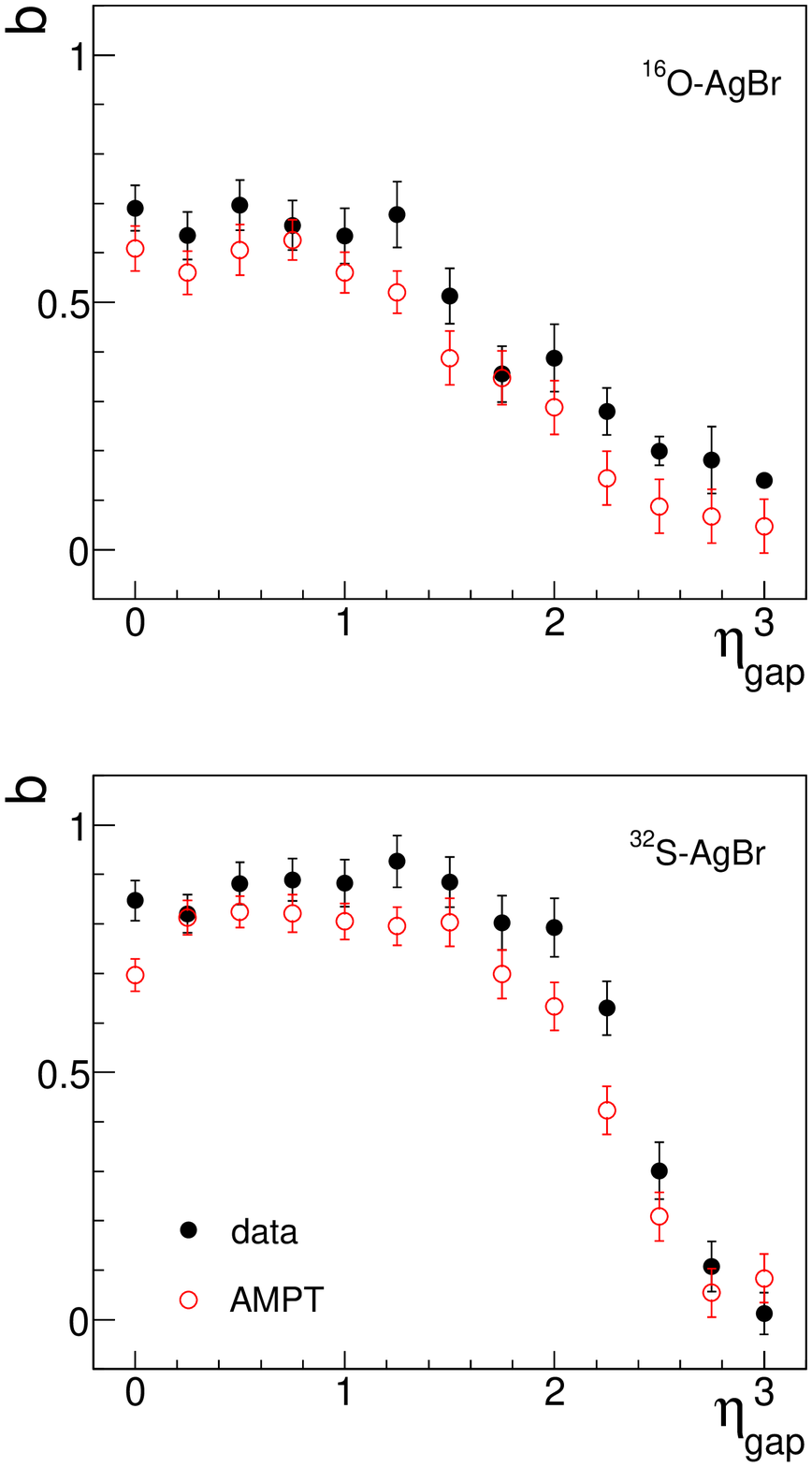,width=10cm,height=16cm}}\end{center}
\caption[1.]{{\sf Dependence of correlation strength, b on separation gap between two symmetric pseudorapidity windows, \(\eta_{gap}\) for various data sets. }}
\label{lindat}
\end{figure}

\end{document}